\documentclass{article}
\usepackage[T1]{fontenc}
\usepackage{spconf,amsmath,amssymb,graphicx,booktabs,multirow}

\newcommand{\etaest}{\widehat{\eta}}
\newcommand{\method}{PhysicsSliders}
\newcommand{\ours}{PXS}

\title{PhysicsSliders: Learning Physical Controls\\
for Audio Foley Generation}

\name{\shortstack{Peiwei Chang\textsuperscript{1}, Yongqiang Chen\textsuperscript{1},  Yuelan Cheng\textsuperscript{1}, Liting Gao\textsuperscript{1}, Yishun Lu\textsuperscript{2},\\ Li Zhang\textsuperscript{3}, Mark D. Plumbley\textsuperscript{4},
Philip J.B. Jackson\textsuperscript{1}, Wenwu Wang\textsuperscript{1}}}
\address{\textsuperscript{1}Centre for Vision, Speech and Signal Processing, University of Surrey, UK\\
\textsuperscript{2}Department of Engineering Science, University of Oxford, UK\\
\textsuperscript{3}Department of Computer Science, Royal Holloway, University of London, UK\\
\textsuperscript{4}Department of Informatics, King's College London, UK}

\begin{document}
\ninept
\maketitle

\begin{abstract}
Physical properties provide intuitive controls for sounds made by interacting objects. The challenge is to make a generative model respond predictably to these controls, both individually and in combination. We present \method\ (PXS), trained on simulated contact sounds with 15 material, geometry and contact parameters. Two separately trained variants accept direct parameter inputs or numerical captions. PXS controls hardness, size, speed and damping, supports room descriptions and produces spectrally distinct outputs at fixed targets. When controlling damping and size together, specifying thickness improves frequency accuracy and joint target delivery. Measurements from real recordings define effective-damping targets, which simulator inversion maps to generation inputs without real-audio fine-tuning. Across 18 impact classes from five real-audio datasets, PXS-Physical follows these targets with mean grouped Spearman correlation $\rho=0.976$ and class-median error 0.094 dex (a factor of 1.24).
\end{abstract}


\section{Introduction}
\label{sec:intro}

Contact sounds convey what objects are made of, how they move and how they interact. Recreating these sounds for media is part of Foley production. Controllable audio generation must go beyond event labels to respond predictably to changes in objects and interactions. For games, virtual environments and animation, this could adapt sounds to changing scenes without selecting a recording for every event. Editors could explore alternatives for the same physical setting or change room acoustics. Material, geometry and excitation provide natural controls because they shape attack, resonances and decay.

Existing audio-generation methods offer complementary forms of control. Text prompts specify sound events and scenes~\cite{evans2024stable,hung2026tangoflux}. Temporal-event conditioning controls event timing~\cite{chung2024tfoley}. Acoustic-feature conditioning follows loudness, brightness and pitch trajectories~\cite{floresgarcia2025sketch}, while latent-space guidance targets intensity, pitch and beats~\cite{novack2026latchs}. Continuous semantic controls further support attribute editing and content preservation~\cite{ezra2025freesliders}. These approaches give users expressive control over the acoustic result, but an edit to an object's size or contact hardness must be expressed through those acoustic or semantic coordinates. In contrast, physics-aware methods connect sound to objects and interactions through modal priors~\cite{su2023physics}, geometry and strike position~\cite{wang2025sonicgauss}, or physical-state descriptions~\cite{yang2026fillgauss} and estimated mass and velocity~\cite{oh2026pavas}. Simulation-trained explosion synthesis also demonstrates that neural models can inherit controls from a physically inspired synthesiser~\cite{zong2025learning}. Single-factor video-to-audio benchmarks further assess physical responses under controlled interventions~\cite{li2026flatsounds}. These advances establish physical conditioning and directional responses. Target-driven audio Foley generation additionally requires meeting specified acoustic values, including two at once. This calls for evaluating absolute target accuracy alongside response direction, and identifying which object properties must be supplied for joint control.

We introduce \method\ (PXS) with direct-parameter (PXS-Physical) and numerical-caption (PXS-Text) interfaces. Both share a flow-matching architecture~\cite{lipman2023flow} but train separately on simulation. Our contributions are: \textbf{(1) Measurement-based generation:} we map effective-damping targets from five real-audio datasets to generation inputs through simulator inversion, and demonstrate target delivery without real-audio fine-tuning. \textbf{(2) Joint target control:} we evaluate simultaneous damping and frequency targets, then isolate the role of thickness information by masking it in the same generator. This distinguishes directional response from joint target accuracy. \textbf{(3) Physical and linguistic interfaces:} we learn hardness, size, speed and damping responses that transfer to held-out materials, combine parameter inputs with room descriptions, and produce spectrally distinct outputs at fixed targets.

\begin{figure*}[t]
  \centering
  \includegraphics[width=0.99\textwidth]{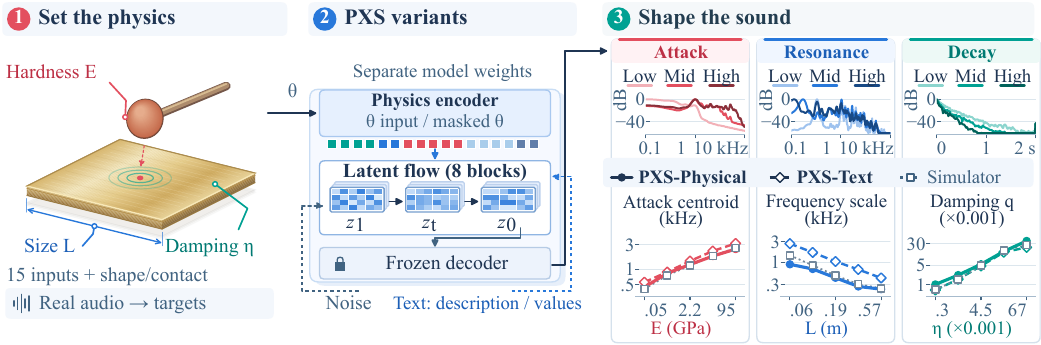}
\caption{PXS variants train separately with a shared architecture and frozen decoder. Paired labels show PXS-Physical first. Objects and latents are schematic. Upper-right brass impacts vary one input at fixed sampling noise. Spectra (attack 30 ms, resonance 500 ms) and RMS envelopes are peak-normalised. Low/mid/high use $E=\{0.05,2.24,100\}$ GPa, $L=\{0.126,0.200,0.316\}$ m and $\eta=\{1.71,4.30,10.8\}\times10^{-3}$. Lower curves show five-level impact sweeps with paired noise. PXS curves geometrically average valid-reading medians (32 samples) across three held-out materials and three runs. Simulator curves average over materials. $q$ is measured effective damping.}
  \label{fig:overview}
\end{figure*}

\section{Method}
\label{sec:method}

PXS-Physical accepts parameters with or without text; PXS-Text accepts text alone (Fig.~\ref{fig:overview}).

\textit{Physical supervision.} Online modal simulation generates 2-second, 44.1-kHz plate and bar sounds from Hertzian impacts and random surface excitation (45\,\% hit, 15\,\% bounce, 25\,\% scrape, 15\,\% roll). The 15-dimensional parameter vector $\theta$ describes material and damping (wave speed $c_L$, density, source loss $\eta$, external damping multiplier $\kappa$), geometry (size $L$, thickness ratio $h/L$), excitation (effective contact modulus $E$, mass, radius, speed $v$, position), and motion (restitution, roughness, normal force, rolling radius). For fixed shape, mode $k$ has frequency $f_k\propto c_L(h/L)/L$ and amplitude-decay rate $d_k=\pi f_k\eta\kappa+1.5\,\mathrm{s}^{-1}$. $E$ controls hardness; $\kappa$ adds damping such as hand muting. Room acoustics, microphone response, noise and RMS gain are randomised. Room 60-dB decay time $T_{60}$ is log-uniform over 0.08--0.8 s. PXS-Physical captions describe the same simulated room as dry below 0.12 s or reverberant above 0.45 s, with no room phrase between these values. Plate aspect ratio is uniform on $[1.15,1.6]$ but is not an input. Size, thickness ratio and hardness are log-uniform. Loss sampling mixes material-specific draws (65\,\%) with full-range coverage (35\,\%), $\eta\in[2.82\times10^{-4},7.10\times10^{-2}]$. Speed is log-uniform over 0.2--5 m/s for impacts and 0.03--1 m/s for sliding; $\kappa=1$ in 25\,\% of draws, otherwise log-uniform on $[1,16]$. Training uses steel, glass, ceramic, stone, wood and plastic, $L=0.06$--$0.6$ m and $h/L=0.005$--$0.10$. Held-out aluminium, brass and concrete have wave speeds 5100, 3430 and 3540 m/s. Aluminium overlaps training support; brass and concrete lie between training bands.

\textit{Conditioned audio generation.} An eight-layer, 54.8-million-parameter 1-D flow transformer is trained from scratch using Stable Audio Open (SAO)'s frozen pretrained Oobleck variational autoencoder (VAE)~\cite{evans2024stable}. Parameter tokens use $a_i=(T_i(\theta_i)-\mu_i)/\sigma_i$, standardised by frozen simulation means $\mu_i$ and standard deviations $\sigma_i$. $T_i$ is the natural log (identity for position, restitution and roughness exponent). Features $\gamma(a_i)$ concatenate $a_i$ with sine and cosine terms at $\pi f$, $f\in\{0.25,0.5,1\}$. Binary mask $m_i$ selects a relevant supplied value or learned null token $n_i$. With identity embedding $p_i$ and shared multilayer perceptron $V$, the token and input at latent position $\tau$ are
\begin{equation}
\begin{aligned}
 e_i &= m_i\{V(\gamma(a_i))+p_i\}+(1-m_i)n_i,\\
 H^{(0)}_\tau &= W_z z_{t,\tau}+p_\tau+W_r[\bar e;\,m\odot a;\,m].
\end{aligned}
\label{eq:physics}
\end{equation}
Shape and contact type add tokens $e_{16},e_{17}$. A learned map $P$ pools all 17: $\bar e=P(\frac1{17}\sum_{i=1}^{17}e_i)$. Here $t\in[0,1]$ is flow time, $W_z,W_r$ are affine maps, $p_\tau$ embeds position, $[\,;\,]$ concatenates and $\odot$ multiplies elementwise. Physics tokens and frozen FLAN-T5-large~\cite{chung2024flan} text embeddings enter cross-attention; pooled physics and time embeddings modulate blocks. Flow matching (FM) learns conditional velocity $v_\phi$ from audio latent $z_0$ to Gaussian noise ($\phi$: weights, $\mathcal C$: conditions):
\begin{equation}
\begin{aligned}
 z_t&=(1-t)z_0+t\epsilon,\quad \epsilon\sim\mathcal N(0,I),\\
 \mathcal L_{\rm FM}&=\mathbb E\lVert v_\phi(z_t,t,\mathcal C)-(\epsilon-z_0)\rVert_2^2.
\end{aligned}
 \label{eq:fm}
\end{equation}
Each variant uses three independent AdamW runs (50,000 updates, batch 48, learning rate $10^{-4}$), evaluated with final exponential moving average weights. During training, PXS-Physical drops text with probability 15\,\%, physics with 14.5\,\% (both: 6\,\%), and individual coordinates with 10\,\%. Physics dropout includes shape and contact type. PXS-Text always masks physics and drops text with 15\,\% probability. Sampling uses 50 Euler steps from Gaussian noise with classifier-free guidance (CFG)~\cite{ho2022cfg} of 3, then the frozen decoder.

\begin{table*}[t]
\centering
\caption{Effective-damping control. (a) Tracking ($n$: classes; error: dex).
(b) Impact requests with two or more of four outputs on target (\%).}
\label{tab:main}
\begin{minipage}[t]{0.53\textwidth}
\centering\small\setlength{\tabcolsep}{3pt}
\textbf{(a) Physical and numeric-text control}\par\vspace{3pt}
\begin{tabular}{@{}lrrrrrr@{}}
\toprule
 & \multicolumn{3}{c}{Impulsive ($n=18$)} & \multicolumn{3}{c}{Sustained ($n=12$)} \\
\cmidrule(lr){2-4}\cmidrule(l){5-7}
Generator & $\rho\uparrow$ & $g\to1$ & Error$\downarrow$ & $\rho\uparrow$ & $g\to1$ & Error$\downarrow$ \\
\midrule
\textbf{PXS-Physical} & .976 & .77 & .094 & .829 & .19 & .293 \\
\textbf{PXS-Text} & .954 & .78 & .058 & .844 & .27 & .360 \\
\bottomrule
\end{tabular}
\end{minipage}\hfill
\begin{minipage}[t]{0.45\textwidth}
\centering\small\setlength{\tabcolsep}{3pt}
\textbf{(b) Multiple on-target outputs ($n=18$)}\par\vspace{3pt}
\begin{tabular}{@{}lrr@{}}
\toprule
Generator & $1.5\times$ & $2\times$ \\
\midrule
\textbf{PXS-Physical} & 83 & 96 \\
\textbf{PXS-Text} & 85 & 96 \\
Simulator (fixed aspect) & 90 & 100 \\
Simulator (varied aspect) & 92 & 100 \\
\bottomrule
\end{tabular}
\end{minipage}
\end{table*}

\begin{table}[t]
\centering\small\setlength{\tabcolsep}{4pt}
\caption{Encoding ablation: model/simulator slope ratios (two runs; ideal 1).
Smooth: Eq.~\ref{eq:physics}; High-freq.: $f=2^j$, $j=0,\ldots,7$, no scalar.
Direct: value/mask path. Bold: PXS-Physical.}
\label{tab:ablation}
\begin{tabular}{@{}lcrrr@{}}
\toprule
Token features & Direct & Hardness & Size & Damping \\
\midrule
\textbf{Smooth} & \textbf{Yes} & 0.95 & 0.76 & 0.97 \\
High-freq. & Yes & 1.11 & 0.65 & 0.89 \\
Smooth & No & 0.94 & 0.65 & 0.97 \\
High-freq. & No & 0.20 & 0.19 & 0.25 \\
\bottomrule
\end{tabular}
\end{table}

\begin{table}[t]
\centering\small\setlength{\tabcolsep}{2.3pt}
\caption{Two-second zero-shot grouped rank.
Missing: incomplete class responses (scored zero).
$p_H$: one-sided class-paired Wilcoxon signed-rank tests (PXS greater),
Holm-corrected across eight comparisons.}
\label{tab:external}
\begin{tabular}{@{}lrrrrrr@{}}
\toprule
 & \multicolumn{3}{c}{Impulsive ($n=18$)} & \multicolumn{3}{c}{Sustained ($n=12$)} \\
\cmidrule(lr){2-4}\cmidrule(l){5-7}
Generator & $\rho\uparrow$ & Missing & $p_H$ & $\rho\uparrow$ & Missing & $p_H$ \\
\midrule
\textbf{PXS-Physical} & .976 & 0 & -- & .829 & 0 & -- \\
\midrule
SAO numeric & .022 & 3 & $<.001$ & -.108 & 2 & .001 \\
SAO ordinal & .278 & 5 & .002 & .033 & 3 & .010 \\
TangoFlux numeric & -.028 & 17 & $<.001$ & .250 & 1 & .026 \\
TangoFlux ordinal & .100 & 16 & $<.001$ & .233 & 1 & .026 \\
\bottomrule
\end{tabular}
\end{table}

\textit{From measurements to requests.} Users set $\theta$ or request effective damping $q=\etaest=h_c(x)$ from recording $x$, with readout $h_c$ for excitation $c$. For impacts, Hilbert envelopes track resonance amplitudes; sustained excitation masks free decay, so unbiased autocorrelation envelopes measure decay with lag. Peak detection uses 8192-sample Hann spectra, onset-aligned for impacts or Welch-averaged with 50\,\% overlap for sustained sounds. We retain up to six impact or eight sustained peaks within 45 or 20 dB of the spectral maximum, respectively, excluding peaks below 150 Hz. Rectangular Fourier masks isolate each resonance. Impact half-widths $b_0=0.015f_k+4$ Hz expand to $\max(b_0,5d_k^{(1)}/\pi)$ Hz using initial decay-rate estimate $d_k^{(1)}$. Invalid refits retain the first estimate. Fits start 10 ms after the envelope peak, requiring 20-dB headroom initially and 15 dB for refits above the final 20\,\% envelope median. They stop at a 30-dB fall, 6 dB above this floor or 1.5 s. Sustained bands have half-width $\max(0.05f_k,25)$ Hz. After discarding the first 10\,\% of audio, autocorrelation fits start at $\max(2/f_k,0.002)$ s and stop below 0.15 normalised amplitude or after one quarter of the remaining duration. Valid fits last at least 10 ms with $d_k>0$. Their median $d_k/(\pi f_k)$ defines $q$, missing if no peak is valid.

Excitation, rooms, modal frequencies and $\kappa$ affect $q$, so it need not equal source loss $\eta$. For object/contact template $\mathcal T$, simulator $S$ maps between them using 24 log-spaced training-range losses $\mathcal E$ and three excitation/recording realisations $\nu_b$ per loss:
\begin{equation}
 \eta^*(q;\mathcal T)=\underset{\eta\in\mathcal E}{\arg\min}
 \left|\log_{10}\frac{\operatorname{med}_{b=1}^{3}h_c(S(\eta;\mathcal T,\nu_b))}{q}\right|.
 \label{eq:inverse}
\end{equation}
Here $\operatorname{med}$ takes the median of valid readings. Template loss $\eta^*$ specifies input $\theta_i^*$ for target $i$, with draws $y_{id}$ indexed by $d$. PXS-Text receives $\eta^*$ in captions. Templates use 0.30-m plates ($h/L=0.008$, $\kappa=1$), class-assigned materials (wood if unspecified), and hits for impulsive targets or scrapes otherwise.

\section{Experiments}
\label{sec:results}

We assess real-audio target accuracy (Table~\ref{tab:main}) and joint control (Fig.~\ref{fig:dual}). Evaluation includes 10 ESC-50~\cite{piczak2015esc50}, 4 FSD50K~\cite{fonseca2022fsd50k}, 6 AudioSet~\cite{gemmeke2017audioset}, 5 NSynth~\cite{engel2017nsynth} and 5 RealImpact~\cite{clarke2023realimpact} classes.

\textit{Targets and grouping.} We retain 18 impulsive and 12 sustained classes after excluding two with insufficient range and two AudioSet/FSD50K duplicates. After trimming endpoints separated by over $2\delta$ while retaining at least eight recordings, each class supplies eight evenly indexed targets from sorted valid $\log_{10}q$. Ordering and gain require at least three groups, each spanning at most $\delta$ from its first log-value. Dex denotes base-10 log units: 0.10 dex is a factor of 1.26. Calibrated material- and excitation-specific thresholds span 0.065--0.491 dex and set ranking resolution, not target tolerance. Impact thresholds average within-strike microphone-pair log-difference 95th percentiles from RealImpact. Sustained thresholds use Arni recordings~\cite{prawda2022arni}, taking the root-sum-of-squares of percentiles within and across configurations. We evaluate all 240 requests, including those outside simulator support.

\textit{Evaluation metrics.} Each target's median $\tilde q_i$ requires at least three of four valid outputs. Gain $g$ is the slope of log group-mean predictions against log group-mean targets, fitted with an intercept (ideal 1). Error is $|\log_{10}(\tilde q_i/q_i)|$, and Spearman $\rho$ measures group ordering. Flat or incomplete class responses score zero correlation and gain. We average median target errors and gains over three runs, then report class medians and mean rank. Selection and qualification use no model outputs. At tolerance $\lambda\in\{1.5,2\}$, requests succeed if at least two of four outputs have $h_c(y_{id})\in[q_i/\lambda,\lambda q_i]$. Invalid readings fail. We average success over requests and training runs or simulator noise banks, then take class medians.

\textit{Measurement validation.} Of 36 dry impulses with known decay, 32 yield valid fits and 29 are within $1.5\times$. An alternative readout is $\ln(1000)/(\pi f_{\rm scale}T_{60})$, using Schroeder decay time $T_{60}$~\cite{schroeder1965decay} and frequency scale $f_{\rm scale}$, the geometric mean of up to three lowest decay-validated resonances. Both readouts are valid for all 144 PXS-Physical impact requests (three runs) and independent simulator templates. Ungrouped correlation and $1.5\times$ request-median agreement are 0.941 and 78.9\,\% for $h_c$, versus 0.986 and 40.7\,\% for the alternative. Ordering is stable, but absolute agreement depends on the decay definition.

\subsection{Target delivery and physical responses}
\label{sec:families}
\label{sec:control}

Both learned interfaces track damping targets from real impacts, with PXS-Physical gain 0.77 (95\,\% class-bootstrap confidence interval $[0.65,0.86]$). Classes have 3--7 groups (median 4). Without grouping, correlations remain 0.920 for PXS-Physical and 0.911 for PXS-Text.
\label{sec:response}
\label{sec:floor}
By contrast, sustained responses cover less of the target range. In a paired scrape sweep, PXS-Physical's effective-damping floor is 12.7 times the simulator value. Impacts show no such floor.

\textit{Sampling at fixed targets.} For 144 impact requests (Table~\ref{tab:main}b), we vary only latent noise, fixing parameters and masking text for PXS-Physical, or fixing the caption for PXS-Text. Simulator references share parameters and recording seeds, with aspect fixed at 1.32 or uniform on $[1.15,1.6]$.

Onset-aligned, unit-RMS audio is compared by $D(x,y)=\sqrt{\operatorname{mean}_{\mathcal M}[(A_x-A_y)^2]}$, where $A_x=10\log_{10}(P_x+10^{-8})$ uses mel power $P_x$ from 64 Slaney bands (20 Hz--16 kHz), 2048-sample Hann windows and 512-sample hops at 44.1 kHz. Before selecting on-target pairs, shared mask $\mathcal M$ retains bins within 40 dB of the 12-output mean-power map's peak (four per arm). Pair distances are averaged within each request and run. PXS-Physical and both simulator arms yield $1.5\times$-accurate pairs in 323 of 432 request--run evaluations. Class-median distances are 10.27 dB for PXS-Physical, 7.15 dB for fixed-aspect simulation and 15.03 dB for varied-aspect simulation.

\begin{table}[t]
\centering\small
\caption{Metrics: VGGish Fr\'echet audio distance (FAD)~\cite{kilgour2019fad};
Kullback--Leibler (KL) divergence of pretrained audio neural networks
(PANNs)~\cite{kong2020panns} softmax outputs (reference to generated, nats);
contrastive language--audio pretraining (CLAP)~\cite{wu2023clap}
audio--text cosine for material-and-strike descriptions.
References are 16 independent renders per $\theta^*$ or real target recordings.
SAO captions specify material, plate and strike.}
\label{tab:quality}
\begingroup
\setlength{\tabcolsep}{3pt}
\begin{tabular}{@{}lrrrr@{}}
\toprule
 & \multicolumn{2}{c}{FAD $\downarrow$} & CLAP $\uparrow$ & KL$_{r\Vert g}$ $\downarrow$ \\
\cmidrule(lr){2-3}
Generator & Sim. ref. & Real ref. & Semantic & Sim. ref. \\
\midrule
Simulator & 0.13 & 4.23 & 0.486 & 0.070 \\
Simulator + VAE & 1.13 & 3.62 & 0.456 & 0.436 \\
\textbf{PXS-Physical} & 4.05 & 3.43 & 0.433 & 0.709 \\
\textbf{PXS-Text} & 2.67 & 2.82 & 0.437 & 0.925 \\
SAO $q$ & 4.16 & 2.37 & 0.404 & 2.896 \\
SAO $\eta^*$ & 4.10 & 2.40 & 0.415 & 2.865 \\
\bottomrule
\end{tabular}
\endgroup
\end{table}

\textit{Physical responses.} We vary $E,L,\eta,v,\kappa$ individually over five levels on hits and bounces across the three held-out materials (32 paired-noise outputs per setting and run). Log-spaced speed and $\kappa$ span 0.21--4.75 m/s and 1.05--15.2, respectively. Model-to-simulator log--log slope ratios measure directional responses (ideal 1). Simulator qualification requires at least two curves with $\geq3$ valid levels, mean direction-corrected Kendall $\tau\geq0.5$, mean log--log fit $R^2\geq0.8$ and absolute mean slope $\geq0.02$. Hardness and speed use attack centroid (first 30 ms); size uses $f_{\rm scale}$, $\kappa$ uses $T_{60}$ and $\eta$ uses $h_c$. PXS-Physical's speed and $\kappa$ ratios span 0.72--1.41 and 0.89--1.04 (five and four of six material--contact combinations qualify, respectively). With matched training and inputs, either smooth encoding or the direct path preserves hardness and damping responses; removing both weakens them (Table~\ref{tab:ablation}).

\subsection{Conditioning interfaces}
\label{sec:interfaces}

\textit{Trained text.} PXS-Text learns from captions such as ``a glass plate of size 0.30 meters struck at 1.00 meters per second with a striker of hardness 10.0 gigapascal, material loss factor 0.0006''.

\textit{Zero-shot generators.} External numeric prompts use ``\{event\}, material loss factor \{q\}'', inserting raw $q$ without inversion and preserving distinct rounded target values. Ordinal prompts map ordered target groups to evenly spaced, rounded indices among ``ringing and resonant'', ``lightly damped'', ``damped'', ``heavily damped'' and ``completely muted''. SAO 1.0 uses 100 steps, CFG 7 and negative prompt ``low quality''. TangoFlux uses 50 steps and CFG 4.5. Both request 2 s with four shared seeds. We retain the first 2 s in mono, matching corpus bandwidth and $-30$ dB RMS. PXS-Physical leads eight failure-aware ordering comparisons under this protocol (Table~\ref{tab:external}).

We test timing sensitivity with predefined 2-second energy windows from 10-second outputs. With matched plate captions, SAO completes all 18 impact classes (correlations -0.085 for $q$ and -0.158 for $\eta^*$). TangoFlux completes 14/18 numeric and 10/18 ordinal classes, with complete-class mean correlations -0.534 and 0.535.

SonicGauss~\cite{wang2025sonicgauss} receives geometry and contact coordinates, while PXS-Physical uses reference damping and proxy plates. We compare four energy-selected 2-s outputs at each of two validation positions on four ObjectFolder-Real objects~\cite{gao2023objectfolder} (one PXS run). Errors average within positions, then objects, using 30 valid SonicGauss and 32 PXS outputs: 0.440 versus 0.171 dex for damping and 0.366 versus 0.520 dex for frequency.

\label{sec:acoustics}
\textit{Audio distribution and semantic alignment.} Table~\ref{tab:quality} compares 576 outputs per model at matched duration, RMS and bandwidth, using three-run PXS averages and fixed-aspect simulation. With four-output medians, VAE reconstruction raises simulator class-median error to real targets from 0.032 to 0.047 dex but lowers real-reference FAD.

\begin{figure}[t]
\centering
\includegraphics[width=\columnwidth]{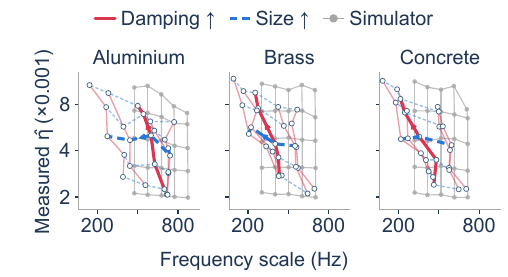}
\caption{Damping--size grid (log axes). Open points geometrically average PXS-Physical four-output medians over three runs. Grey points are eight-output simulator medians. Arrows increase damping (red) or size (blue). Bold lines fix the other control at its midpoint; faint lines show full grids.}
\label{fig:dual}
\end{figure}

\subsection{Joint control}
\label{sec:joint}

A $5\times5$ grid (Fig.~\ref{fig:dual}) spans $\eta$ 1.7--10.8$\times10^{-3}$ and $L$ 0.13--0.32 m on the three held-out materials. Before generation, all materials pass simulator checks: $\geq6$ valid readings from eight outputs per cell, strictly monotone axes spanning at least half the requested log range, and $\geq0.05$ dex separation in $(\log_{10}q,\log_{10}f_{\rm scale})$. Training-format captions specify material, $L,v,E,\eta$. Physical inputs add material constants, $h/L=0.008$ and contact geometry. Both follow requested directions (slope ratios 0.82--0.98).

Joint success requires at least two of four outputs within $1.5\times$ of both eight-output simulator medians. PXS-Physical leads PXS-Text in every run: 68.0\,\% versus 15.6\,\% over 225 setting--run evaluations. Frequency error is lower (0.126 versus 0.372 dex), while damping errors are similar (0.055 versus 0.053 dex). Cross-talk slopes (dex/dex) for PXS-Physical and PXS-Text are -0.26 and -0.33 for frequency versus loss, and +0.34 and +0.26 for damping versus size (simulator: zero and +0.08).

Masking only $h/L$ at fixed weights and noise lowers success in every run (27.6\,\% overall), despite a size response of 0.86. Errors are 0.245 dex for frequency and 0.058 dex for damping. With 32-output simulator medians, supplying thickness raises $1.5\times$ joint success from 51.1\,\% to 80.0\,\% and from 57.8\,\% to 95.6\,\% at $h/L=0.016$ and $0.032$ (corner and centre settings), with gains in all three runs.

\textit{Room text and size.} PXS-Physical combines five sizes with dry, reverberant or no room text for brass and concrete ($\eta=0.04$, 360 shared-noise impacts). Reverberant versus dry text raises onset-aligned tail-to-attack energy (80--600 versus 0--80 ms) in all 30 comparisons, with context medians of 5.75--10.10 dB. All 18 size slopes remain negative, with frequency-scale shifts of +0.097--0.240 dex.

\section{Conclusion}
\label{sec:discussion}

Simulation-trained \ours\ provides direct-parameter and numerical-caption interfaces that deliver effective-damping targets from real impacts without real-audio fine-tuning. Joint tests distinguish directional response from meeting two acoustic targets: specifying thickness improves frequency accuracy and joint delivery. Parameter inputs combine with room descriptions, while repeated sampling produces spectrally distinct sounds at fixed targets. Future work targets sustained range compression, cross-talk and perceptual evaluation.

\label{sec:bodyend}
\clearpage

\section{Acknowledgements}\label{sec:acknowledgements}

This work was supported by the UKRI AI Centre for Doctoral Training in AI for Digital Media Inclusion [grant number EP/Y030915/1] and by the Engineering and Physical Sciences Research Council (EPSRC) [grant number EP/Y028805/1]. For the purpose of open access, the authors have applied a Creative Commons Attribution (CC BY) licence to any Author Accepted Manuscript version arising from this submission.

\end{document}